# Dramatic enhancement of supercontinuum generation in elliptically-polarized laser filaments

Shermineh Rostami[1], Michael Chini[2], Khan Lim[2], John P. Palastro[3], Magali Durand[2], Jean-Claude Diels[1], Ladan Arissian[1], Matthieu Baudelet[2] and Martin Richardson[2]

[1]Center for High Tech Materials, University of New Mexico, Albuquerque NM 87106

[2]Townes Laser Institute, CREOL – The College of Optics and Photonics, University of Central Florida, Orlando FL 32816, USA

[3] Plasma Physics Division, Naval Research Laboratory, Washington, DC 20375-5346, USA

**Broadband laser sources based on supercontinuum generation by femtosecond laser filamentation[1,2] have enabled applications from stand-off sensing and spectroscopy[3] to the generation and self-compression of high-energy few-cycle pulses[4,5]. Filamentation relies on the dynamic balance between self-focusing and plasma defocusing[6], mediated by the Kerr nonlinearity and multiphoton ionization respectively. The filament properties, including the supercontinuum generation, are therefore highly sensitive to the properties of both the laser source and the propagation medium. Here, we report the anomalous spectral broadening of the supercontinuum for filamentation in molecular gases, which is observed for specific elliptical polarization states of the input laser pulse. The resulting spectrum is accompanied by a modification of the supercontinuum polarization state and a lengthening of the filament plasma column. Numerical simulations confirm that the anomalous behavior originates from the delayed rotational response of the medium, demonstrating a new parameter to control supercontinuum generation.**

Supercontinuum generation (SCG) resulting from laser filamentation in gases was first observed 20 years ago[1,7] and has since attracted the interest of scientists in diverse fields of research. It has proven critical to both the understanding of laser-matter interactions underlying filamentation and the generation of secondary sources for sensing (LIDAR) and countermeasure applications, for instance. Despite this long history, there remains a need for more complete understanding of the mechanisms underlying filament-induced SCG. Numerous factors appear to play a significant role in filament SCG, including the chirp[8,9] and polarization[10,11] of the input laser pulse, as well as control of the focusing process before filamentation by preparation in the medium[12,13] or the focusing conditions[6].

In this letter, the effect of the laser polarization on filamentation and the consequent SCG is studied. Despite clear understanding of SCG generated by linear polarization, there remains controversy when it comes to circular polarization. Different studies (under various experimental or simulated conditions) have reported circularly polarized filaments to be more or less efficient for supercontinuum generation[11, 14-16], as well as plasma formation[10, 16]. Furthermore, the polarization has been observed to change during the filamentation of elliptically polarized pulses in air[17], along with debates on the stability of light ellipticity during and following filamentation[18-20].

Here, we present a combined experimental and theoretical study of the supercontinuum spectrum created by filamentation of elliptically polarized laser pulses. Our measurements show an enhancement of the supercontinuum bandwidth generated during the filamentation of pulses with specific polarization ellipticities in molecular gases. The experimental results and accompanying simulations suggest that rotational dynamics of the diatomic molecules play an essential role in the filamentation-induced SCG, which can be controlled with light ellipticity.

We measure the spectrum and polarization ellipticity of the supercontinuum generated through filamentation of a femtosecond laser pulse (50 *fs*, 2.8 mJ)[21] in different gases and for different prepared input polarization ellipticities. The input ellipticity is controlled by rotating a broadband quarter wave plate (QWP) before filamentation is aided by low numerical aperture focusing[6] (see Fig. 1a). In nitrogen, we observe a dramatic anomalous broadening of the supercontinuum spectrum (Fig. 1b) for quarter wave plate angles of 39º and 51º degrees (input ellipticity $\varepsilon_{in} \approx 0.7$). In oxygen (see Supplementary Fig. 1) the broadest spectra appear for QWP angles of 43 and 47 degrees ($\varepsilon_{in} \approx 0.8$). The ellipticity of the beam (ratio of the minor to major axis of the polarization ellipse) after filamentation in nitrogen is plotted in Figure 1c as a function of the QWP angle. It is clear that propagation of the filament significantly modifies the polarization at the same angles that produce enhanced spectral broadening. Figures 1d and 1e show no anomalous polarization dependent features for the same measurements performed in argon, which has a comparable ionization potential to nitrogen (15.76 eV and 15.58 eV, respectively).

Comparing the supercontinuum spectrum generated by single filaments in atomic (argon and krypton) and molecular gases (nitrogen, oxygen and air) (see Fig. 1 and Supplementary Fig. 1), the observed anomalous behavior is evidently unique to filament-induced SCG in molecular gases. We observe a broader white light spectrum for linear polarization than for circular polarization, with the bandwidth decreasing monotonically with increasing ellipticity in atomic gases. In molecular gases however, for a specific input ellipticity the spectrum extends well beyond the shortest wavelengths observed with linear input polarization, while the output ellipticity exhibits a remarkable modification. In every case, the behavior is symmetric about circular polarization, indicating that the broadening is identical for left- and right-handed elliptical polarizations.

Previous experiments with elliptically polarized pulses have also shown unusual modification of the pulse polarization after filamentation in air[17]. Under comparable conditions (60 fs, 5 mJ pulses inducing single filamentation in air) to the experiments presented here, substantial modification of the polarization was observed at similar input ellipticity close to circular polarization. However no unusual behavior was observed for filaments prepared under vacuum focusing conditions[12] (See Supplementary Fig. 2), which shows that the nonlinear focusing (involving the Kerr response of the molecular medium) is necessary for the observation of anomalous behavior in filaments induced by elliptically polarized pulses.

The peculiar behavior described above must be the result of polarization-induced modifications to the filament properties. Indeed, we observe an unexpected lengthening of the filament, which is correlates directly to the spectral broadening and polarization modification in molecular gases. To measure the filament length, transverse plasma emission from $N_2$ is imaged with a camera as the angle of the quarter wave plate is changed. The results shown in Fig. 2, indicate that the collapse position gradually moves away from the lens as the polarization changes from linear to circular. This is consistent with the reduced strength of $\chi^{(3)}$ effects for circular as compared to linear polarization, resulting in a weaker self-focusing and delaying the onset of filamentation[10]. However, for the QWP angle of 39 degrees, the filament is observed to extend further from the focusing lens, indicating that the self-focusing and plasma defocusing are balanced over a longer distance.

It is well known that molecular rotation can play a significant role in filamentation, with modified polarization[22], broader supercontinuum spectra[23] and longer plasma channels[24] observed for filamentation in pre-aligned media. However, determining the role of the rotational response in a single pulse remains challenging for both experimentalists and theorists. Early modeling of nonlinear propagation in molecular media by Close[25], indicated that for elliptical polarization, left- and right-handed circularly polarized components of the field experience different nonlinear indices of refraction. Later modeling by Kolesik *et al.*[16], which included the vector nature of the Kerr nonlinearity (i.e. cross- as well as self-Kerr effects), losses due to ionization, and the delayed nonlinear susceptibility associated with molecular rotation, suggested a more complex behavior of the polarization after filamentation. For an elliptically polarized input pulse, the filamentation process resulted in a more circularly polarized output in the filament core, without indication of a sudden broadening of the supercontinuum or modification of the polarization. However, even this more realistic treatment may significantly underestimate the coupling between different polarization components due to simplified classical modeling of the delayed molecular rotational response.

In this study, theoretical modeling of filamentation in molecular gases was performed by including a self-consistent linear density matrix treatment of the rotational dielectric response of the molecules subjected to arbitrary light polarizations in a nonlinear propagation model (see Methods). Previous simulations carried out under the same model[20], indicate that strong coupling of left- and right-handed circular polarization states arises from off-diagonal terms in the molecular susceptibility tensor, resulting in similar anomalous behavior to what we observe in the experiments. Simulated and experimentally measured supercontinuum spectra resulting from filamentation in nitrogen and oxygen gases are compared in Fig. 3. The strong qualitative agreement between experiments (Fig. 3a, c) and simulations (Fig. 3b, d) carried out under identical conditions, along with the fact that there is no anomalous behavior in atomic gases, indicate that the necessary physics for describing the anomalous behavior is included in the rotational contribution to the molecular susceptibility. Other mechanisms, such as ellipticity dependences in the ionization rates or population of excited states, could be involved in explanation of other ellipticity dependent filamentation processes[26]. However, such effects alone cannot explain the observed enhancement of the supercontinuum spectrum.

In summary, we have demonstrated a dramatic enhancement in the spectral broadening of filament-induced SCG, which can be controlled by tuning the initial ellipticity of the laser pulse undergoing filamentation. This effect, which is observed only for molecular gases, is correlated to a decrease in the ellipticity of the beam after filamentation and an increase in the length of the filament plasma channel. Both experimental and simulated results clearly demonstrate the strong dependence of filamentation on the input laser ellipticity via the delayed rotational response of the molecular medium. This study introduces an additional tool – polarization ellipticity – to control SCG and enhance the white light spectrum, suggesting that further refinements of filamentation models may provide a route to better understanding the dynamic laser-matter interaction underlying filament propagation.

**Methods**

**Supercontinuum generation and diagnostic**

In the experiments, femtosecond laser pulses (pulse duration $\tau_p$ = 50 fs, energy $E$ = 2.8 mJ, wavelength $\lambda$ = 800 nm) from the Multi-Terawatt Femtosecond Laser (MTFL) facility[21] at the University of Central Florida were focused by a lens (focal length $f$ = 3 m), into a transparent gas-filled chamber (length $L$ = 4.5 m). The experimental set-up is shown in Fig. 1a. Elliptically polarized laser pulses, created using a zero order QWP placed just before the focusing lens, were used to produce single (through pressure control) filaments individually in air, nitrogen ($N_2$), oxygen ($O_2$), argon, and krypton. The spectrum of the white light supercontinuum after filamentation was measured after scattering by a diffuser, using spectrometers covering visible ($\lambda$ = 300-740 nm, Ocean Optics HR2000) and near-infrared ($\lambda$ = 660-930 nm, Ocean Optics USB2000) regions of the spectrum. Polarization of the beam after filamentation was determined by fitting the polarization ellipse obtained by measuring the transmitted energy (attenuated using a neutral density filter) through a rotating polarizing cube, for each QWP angle. For both the spectral and polarization measurements, the detectors were placed approximately 2.2 meters after the geometric focus of the lens.

**Simulated spectra**

The laser pulse evolution and supercontinuum generation are simulated using the modified paraxial wave equation in azimuthally symmetric, cylindrical coordinates. The transverse, vector electric field, $\mathbf{E}_F$, is expressed as an envelope $\mathbf{E}$, modulated by phase: $\mathbf{E}_F = \mathbf{E}(r,z,t)e^{-ik\xi} + c.c.$, where the pulse envelope evolves according to:

$$\left[\nabla_\perp^2 + 2\frac{\partial}{\partial z}\left(ik - \frac{\partial}{\partial \xi}\right) - \beta_2 \frac{\partial^2}{\partial \xi^2}\right]\mathbf{E} = 4\pi\left(ik - \frac{\partial}{\partial \xi}\right)^2 \mathbf{P}_{NL}. \qquad (1)$$

In Eq. (1) $k = \omega_0 c^{-1}[1 + \delta\varepsilon(\omega_0)/2]$, $\delta\varepsilon(\omega)$ is the shift in dielectric constant due to linear dispersion, $\xi = v_g t - z$ is the group velocity frame coordinate, $v_g = c[1 - \delta\varepsilon(\omega_0)/2]$, and $\beta_2 / \omega_0 c = (\partial^2 k / \partial \omega^2)|_{\omega=\omega_0} = 20$ fs$^2$/m accounts for group velocity dispersion. The nonlinear polarization density, $\mathbf{P}_{NL} = \mathbf{P}_{elec} + \mathbf{P}_{rot} + \mathbf{P}_{free} + \mathbf{P}_{ioniz}$, includes the instantaneous (Kerr) response, the delayed molecular rotational response, the free electron response, and ionization energy losses respectively[20, 27].

The electronic and rotational polarization densities can be expressed as the product of an effective susceptibility matrix and the electric field envelope: $\mathbf{P}_{elec} = \chi_{elec}\mathbf{E}$ and $\mathbf{P}_{rot} = \chi_{rot}\mathbf{E}$. Using the circularly polarized basis, the effective susceptibility matrix elements are given by

$$(\chi_{el})_{LL} = \frac{1}{6\pi^2}\left(\frac{N_g}{N_{atm}}\right)n_{2,g}\left(|E_L|^2 + 2|E_R|^2\right) \quad (2a)$$

$$(\chi_{el})_{LR} = 0 \quad (2b)$$

$$(\chi_{rot})_{LL} = \frac{N_g(\Delta\alpha)^2}{15\hbar}\sum_j C_j \int_{-\infty}^{\xi}\sin[\omega_{j+2,j}(\xi'-\xi)/c]\left(|E_L|^2 + |E_R|^2\right)d\xi' \quad (2c)$$

$$(\chi_{rot})_{LR} = \frac{2N_g(\Delta\alpha)^2}{5\hbar}\sum_j C_j \int_{-\infty}^{\xi}\sin[\omega_{j+2,j}(\xi'-\xi)/c]E_L E_R^* d\xi', \quad (2d)$$

where

$$C_j = \frac{(j+1)(j+2)}{2j+3}\left(\frac{\rho^0_{j+2,j+2}}{2j+5} - \frac{\rho^0_{j,j}}{2j+1}\right) \quad (3)$$

$N_g$ is the gas density, $N_{atm}$ is the gas density at 1 atm, $n_2$ is the instantaneous Kerr coefficient at 1 atm[28], $\Delta\alpha$ is the difference in molecular polarizabilities parallel and perpendicular to the molecular bond axis[29], $j$ is the total angular momentum quantum number, $\omega_{j+2,j} = \hbar(2j+1)/M$, $M$ is the molecular moment of inertia, $\rho^0_{j,j}$ are the thermal equilibrium density matrix elements and the remaining matrix elements can be obtained via L-R suffix interchange[20]. The values of $n_2$ and $\Delta\alpha$ are taken from Refs. 32 and 33, respectively, and are given in Table 1. Rotational states up to $j = 25$ were included in the sums.

| Gas species | $n_2$ ($10^{-20}$ cm²/W) | $\Delta\alpha$ ($10^{-25}$ cm³) |
|---|---|---|
| N₂ | 7.4 | 9.3 |
| O₂ | 9.5 | 11.4 |

**Table 1 | Nonlinear coefficients used in the simulations.**

The free electron and ionization damping polarization densities are determined by $(ik - \partial_\xi)^2 \mathbf{P}_{free} = (4\pi)^{-1}k_p^2\mathbf{E}$ and $(ik - \partial_\xi)\mathbf{P}_{damp} = -\kappa_{ion}\mathbf{E}$, where:

$$\kappa_{ion} = \frac{1}{2c} U_I \nu_I n_g \frac{1}{|E|^2} \tag{4}$$

is the damping rate, $k_p^2 = 4\pi e^2 n_e / m_e c^2$ is the plasma wavenumber, $n_e$ is the free electron density, $U_I$ is the ionization potential, $\nu_I$ the cycle averaged ionization rate[20, 30], and $\partial_\xi n_e = c^{-1} \nu_I n_g$. Additional details about the response model can be found in Ref. 20.

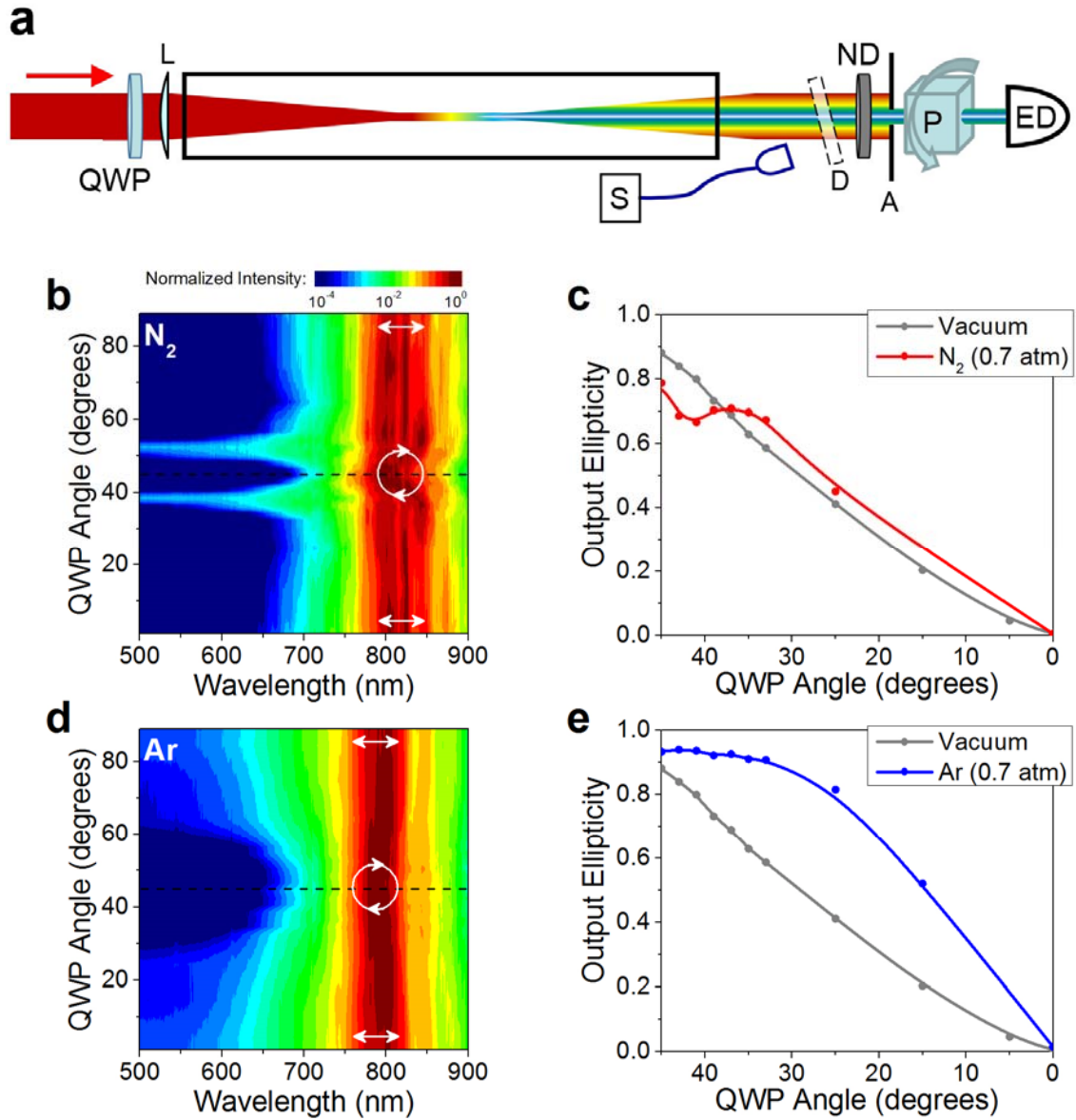

**Fig. 1 | Anomalous behavior in supercontinuum and polarization. a,** Experimental setup to measure the supercontinuum spectrum and polarization ellipticity. The 50 fs, 2.8 mJ pulses at 800 nm are focused by a lens ($f$ = 3 m) into a 4.5 m gas chamber. A quarter wave plate (QWP) is rotated to prepare the beam with different initial elliptical polarization. After attenuation by a neutral density filter (ND), the central part of the beam is selected by an aperture (A). A rotating polarizer cube and an energy detector (ED) are then used to record the polarization ellipse. A fiber-coupled spectrometer (S) records the scattering from a diffuser (D). **b,** Spectral intensity of the supercontinuum spectrum as a function of wavelength and QWP angle measured in nitrogen. White arrows indicate the angles for linear and near-circular polarization. **c,** Ellipticity measured after the filament (red) created in nitrogen and with the chamber evacuated (gray) as a function of the quarter wave plate angle. **d, e** Same measurements as in **b, c** done in argon.

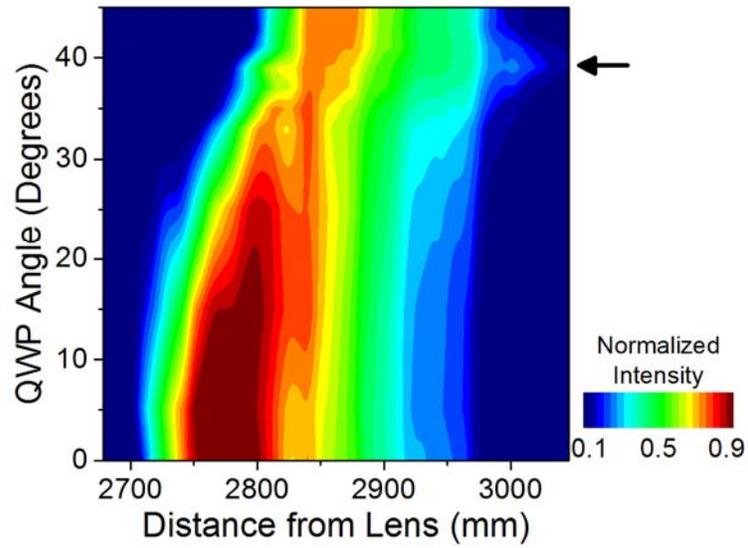

**Fig. 2 | Ellipticity dependence of filament plasma channel.** The intensity of the transverse plasma emission in $N_2$ is plotted for various QWP angles, exhibiting an anomalous extension away from the focal lens when the QWP angle is set to 39 degrees, as indicated by the arrow.

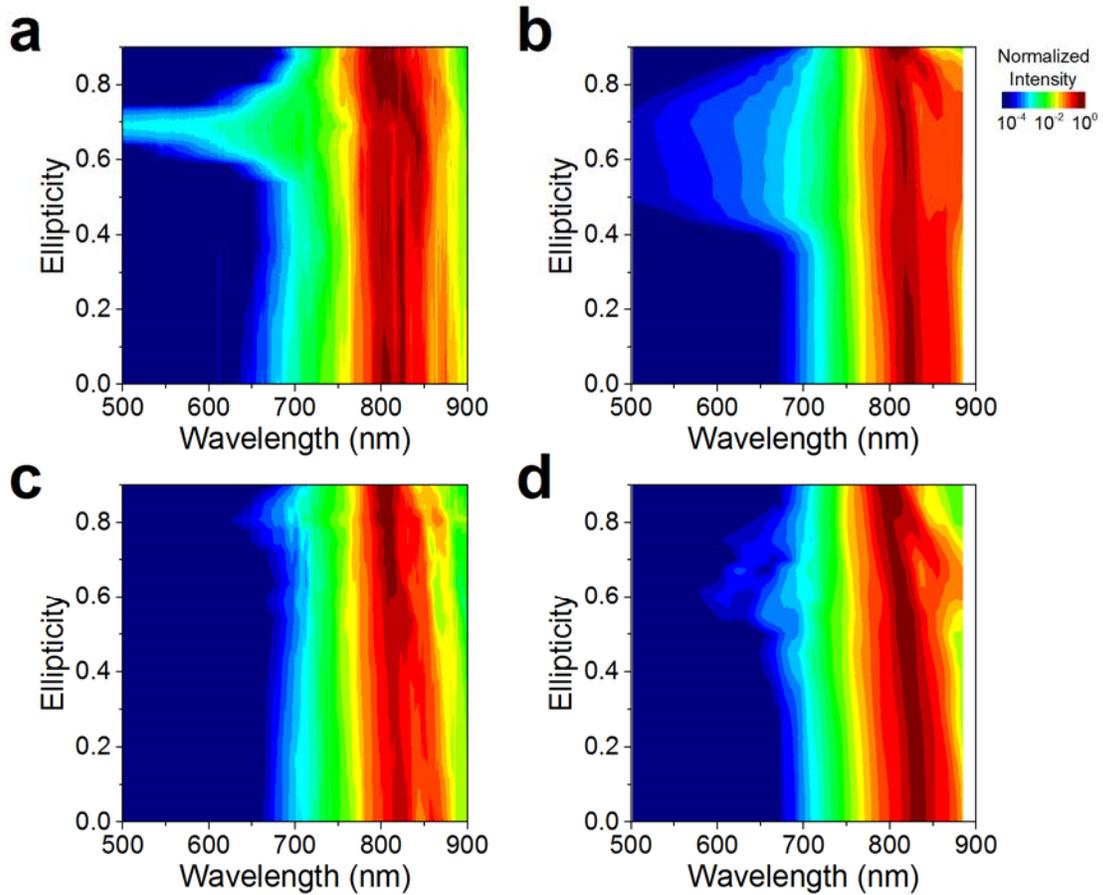

**Fig. 3 | Dependence of supercontinuum spectra on the input ellipticity.** Experimental (**a**, **c**) and simulated (**b**, **d**) results are compared for filamentation in nitrogen (**a**, **b**) and oxygen (**c**, **d**) gases. Experiments and simulations in atomic gases (Ar and Kr), showed no anomalous behavior.

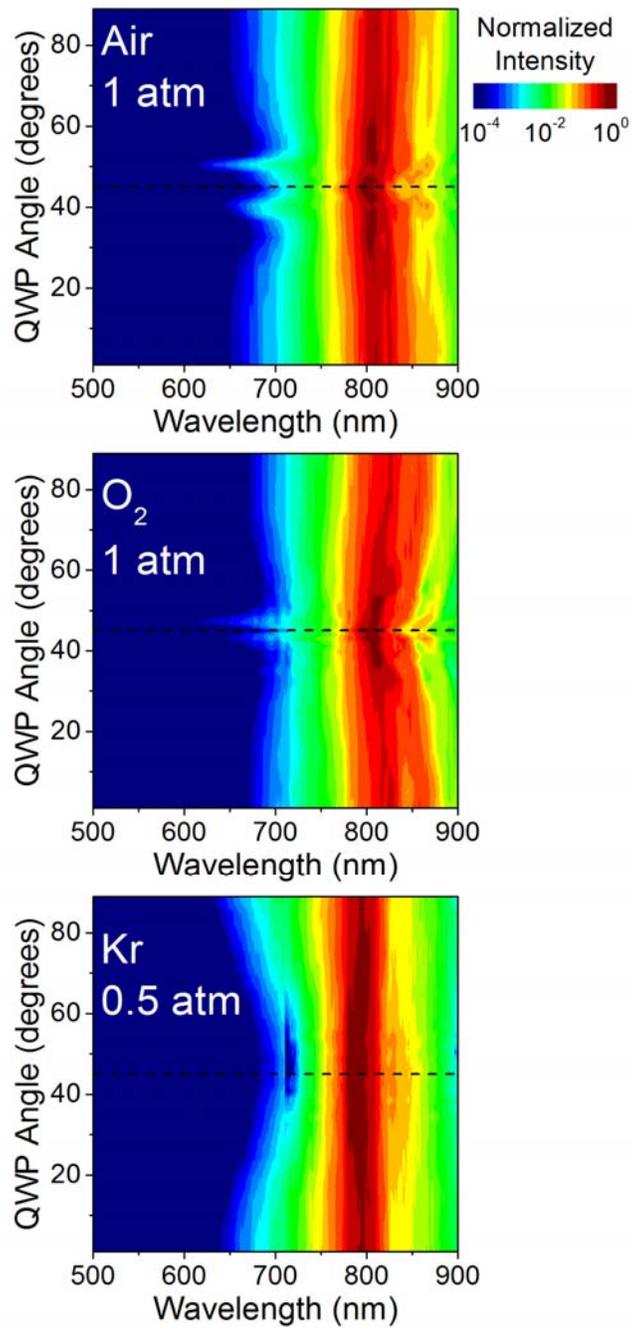

**Supplemental Fig. 1 | Ellipticity-dependent supercontinuum spectrum produced from single filaments in air, oxygen, and krypton gases.**

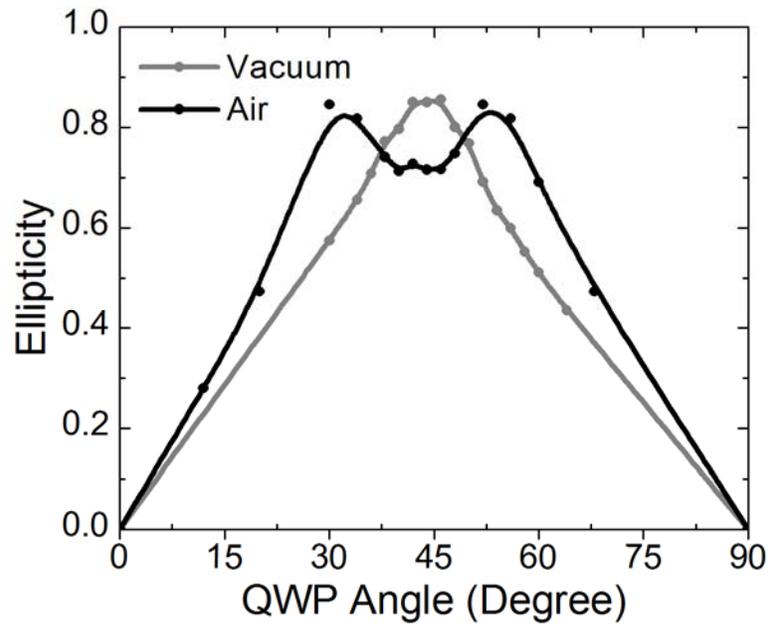

**Supplemental Fig. 2 | Output ellipticity for air filaments prepared by focusing in vacuum (gray circles) and in air (black squares).**